\documentclass{article}
\usepackage{amsmath,amssymb}
\usepackage{epsfig}
\begin{document}
\begin{center}
{\bf\Large Bremsstrahlung Response of Homogeneous Magnetoactive
Plasma on a Gravitational Wave}\\[12pt]
Yu.G.Ignatyev, A.A.Agathonov\\[2pt]
Kazan State Pedagogical University,\\ Mezhlauk str., 1, Kazan
420021, Russia
\end{center}

\begin{abstract}
Numeric model of the bremsstrahlung response of homogeneous
magnetoactive plasma on a gravitational wave with $e_+$
polarization was constructed. Electromagnetic response
dependencies on the plasma and gravitational wave parameters were
determined.
\end{abstract}

\section{Introduction}
The equations of the relativistic magneto\-hydro\-dynamics (RMHD) of a
mag\-ne\-to\-ac\-tive plas\-ma in a gravitational field were formulated in
paper \cite{B1}\footnote{Till 2000 Yu.G. Ignatyev wrote his name as Yu.G.
Ignat'ev.} using the equality requirements for dynamic velocities of plas\-ma
and electromagnetic field\footnote{This requirement is completely equivalent to
the condition of plasma infinite conductivity, see Ref.\cite{B1}.}. These
equations were obtained on the basis of the Einstein and Maxwell equations.
Also the remarkable class of {\it exact solutions} of obtained RMHD equations
was found. It explains movement of a mag\-ne\-to\-ac\-tive locally isotropic
plasma in a field of plane gravi\-ta\-tional wave (PGW). This class was called
{\it gra\-vi\-mag\-ne\-tic shock waves} (GMSW). It describes {\it essentially
nonlinear processes} which do not exist in a linear approximation of the
mag\-ne\-to\-hydro\-dynamics and essentially relativistic processes in terms of
domina\-tion of massless electromagnetic component in a mag\-ne\-to\-ac\-tive
plasma.

It was shown in paper \cite{B2} that the GMSW in pulsars
magnetospheres may be the highly effective detectors of
gravitational waves from neutron stars. Particularly, giant pulses
which sporadically appear in radiation of some pulsars may be the
observation result of transferring energy from a gravitational
wave to the GMSW. Estimations made in \cite{B2}, \cite{B3} allow
to connect giant pulses in radiation of pulsar B0531+21 with
gravitational radiation in the basic mode of oscillations from
this pulsar. In fact, at this moment it is sufficiently difficult
to say about identification of giant pulses as the electromagnetic
display of gravi\-magnetic shock wave evolution in pulsar
magnetosphere and to connect unambiguously these pulses with
pulsars gravi\-ta\-tional radiation. Nevertheless, the idea of
analyzing the influence of gravitational waves from a compact
astro\-phy\-sical object on its own electromagnetic radiation is
highly productive for solving the problems of gravitational waves
detection.

In fact, the main difficulties of gravitational waves detection in
the Earth conditions are:
\begin{enumerate}
\item Negligible amplitude of gravitational waves in the Earth
conditions ($h\lesssim 10^{-19}$) because of significant distance
between relativistic astro\-phy\-sical objects and the Earth.

\item Sporadic nature of events leading to radiation of
gravitational waves inside relativistic astro\-phy\-sical objects
with enough power. This does not allow to connect unambiguously
received signal with a fact of gravitation radia\-tion detection.

\item Impossibility to construct relativistic detectors with
anomalous highly effective parameters for registration of
gravitational waves in con\-di\-tions of the Earth laboratory
(super-strong magnetic fields, highly anisotropic working body of
detector, low level of background noise, etc.).
\end{enumerate}

It is possible to avoid these problems if one can transfer
detector directly close to a relativistic astrophysical object. In
this case one always has prepared electromagnetic signal and there
is no need to convert it in other forms so it allows to do
correlation analysis. If the detector's working body is the
magnetosphere of relativistic astrophysical object, the optimal
for gravitational waves regi\-stra\-tion parameters of detector's
working body will be achieved automatically: super-strong magnetic
fields, ultra\-re\-la\-tivistic equation of state, highly
anisotropy, etc.

The fundamental importance of the GMSW for theory as the direct conversion
effect of gravity-waves energy into electromagnetic energy leads to necessity
of more detailed and comprehensive researches. In Ref.\cite{B4} the strict
proof of the GMSW hydrodynamic theory based on relativistic kinetic theory was
given. In \cite{B1}, \cite{B2}, \cite{B3} was shown that GMSW realizes in
essentially collision\-less non\-equi\-lib\-rium plasma within anomalous strong
magnetic fields. Isotropy of a local plasma electron distribution essentially
violates in the such conditions due to strong brems\-strah\-lung. Therefore, an
anisotropy factor of magnetoactive plasma is highly essential for effectiveness
of GMSW formation mechanism. Hydrodynamic model of GMSW in anisotropic plas\-ma
with adjusted correlation between parallel and per\-pen\-di\-cular components
of plasma pressure was con\-struct\-ed in \cite{B5}. It was based on the
general equations of RMHD. Particularly, in \cite{B5} was con\-sidered the
elementary linear correlation. The research made in \cite{B5} discovered the
strong dependence of GMSW effect upon the plasma anisotropy degree. That fact
led to necessity of constructing the dynamic model of anisotropic magnetoactive
plasma movement in a gravitational radiation field.

Process of a gravitational wave energy pumping over into the
electromagnetic energy is describing with the help of the {\it
energo\-balance equation} intro\-duced in Ref.\cite{B1}-\cite{B3}.
It performs the fact of total momentum conservation law inside the
system ``gravi\-ta\-tional wave + magnetoactive plasma''. In paper
\cite{B6} analytic research of essentially nonlinear equation was
done and some features of the solution were detected. However,
because of software existed in 1998 and other reasons the total
research of GMSW evolution was not done and the parameters of
brems\-strah\-lung response of magnetoactive plasma on
gravitational wave were not obtained. This paper is dedicated to
these problems solution. Here we set the unit system where $c = G
= \hbar = 1$.

\section{Gravimagnetic shock waves}

Let us reproduce the main results of the GMSW theory which are
necessary for the goals of this paper. Let us set the metrics of
vacuum PGW with $e_+$ polarization\footnote{The case of two
polarization states will be considered in next paper.} which
propagates alone the $Ox^1$ axis:
\begin{equation}\label{I.1}
d s^{2}=2 du dv - L^{2}[e^{2 \beta}(dx^{2})^{2} + e^{-2\beta}( dx^{3})^{2}],
\end{equation}
where $\beta(u)$ is an arbitrary function (the PGW amplitude); the
function $L(u)$ (the PGW background factor) obeys an ordinary second
order differential equation\footnote{See, for example,
Ref.\cite{B7}.} ; $ u= \frac{1}{\sqrt{2}}(t - x^{1})$ is the
retarded time and $v= \frac{1}{\sqrt{2}}(t + x^{1})$ is the advanced
time. Let in the absence of PGW ($u\leq 0$) be given a homogeneous
magnetic field directed along the $Ox^2$ axis\footnote{The general
case of a magnetic field arbitrarily directed in the plane $x^1Ox^2$
was considered in the cited papers.}:
\begin{equation}\label{I.2}
H_i(u\leq 0)=\delta^i_2 H_0.
\end{equation}

In general, alternating electromagnetic field which appears in the
presence of gravitational wave has only spacelike magnetic
component $H^i$ in the co\-moving frame of reference moving with
local velocity $v^i$ \cite{B1}:
\begin{equation}\label{H}
H_i=v^k\stackrel{*}{F}_{ki};\qquad
\stackrel{*}{F}_{ki}=\frac{1}{2}\eta_{kilm}F^{lm};
\end{equation}
($F_{ik}$ - Maxwell tensor, $\stackrel{*}{F}_{ki}$ - dual Maxwell
tensor, $\eta_{kilm}$ - discriminant tensor). The electric
component of electromagnetic field in the comoving frame of
reference equals zero \cite{B1}:
\begin{equation}\label{E}
E_i=v^kF_{ki}=0,
\end{equation}
therefore the energy-momentum tensor (EMT) of electromagnetic
field is:
\begin{equation}  \label{I.3}
\stackrel{H}{T}_{ij} = \frac{1}{8\pi} \left( 2 H^2 v_i v_j - 2 H_i
H_j - g_{ij} H^2 \right).
\end{equation}

Squared magnetic field strength is determined as \cite{B1}:
\begin{equation}\label{H^2}
H^2=-(H,H)=\frac{1}{2}F_{lm}F^{lm}.
\end{equation}

Thus, the trace of the EMT of electromagnetic field is equal to
zero:
\begin{equation}\label{T_H}
\stackrel{H}{T}=g^{ij}\stackrel{H}{T}_{ij}\equiv 0.
\end{equation}

Invariants of electromagnetic field comply with the conditions:
\begin{equation}\label{inv_em}
F_{ik}F^{ik}=\stackrel{*}{F}\ \!\!\!
^{ik}\stackrel{*}{F}_{ik}=2H^2>0; \quad
\stackrel{*}{F}_{ik}F^{ik}=0.
\end{equation}

Let further magneto\-active plasma be homo\-geneous but
anisotropic in general in the absence of PGW. In gravitational
field the EMT of anisotropic magneto\-active plasma is \cite{B5}:
\begin{equation}  \label{I.4}
\stackrel{P}{T^{ij}} =  \left(\varepsilon +  p_\perp \right) v^i v^j   -
p_\perp g^{ij}+\left(p_\parallel - p_\perp \right) h^i h^j \,,
\end{equation}
where $h^i=H^i/H$ - spacelike unitary vector of a magnetic field
($(h,h)=-1$), $p_\perp$ è $p_\parallel$ - perpendicular and
parallel components of the plasma's pressure, and according to
(\ref{H}):
\begin{equation}\label{v,h}
(v,h)=0,
\end{equation}

And the EMT (\ref{I.4}) according to virial law complies with the
condition:
\begin{equation}   \label{I.5}
\stackrel{P}{T} = \varepsilon - p_\parallel - 2 p_\perp\geq 0\Leftrightarrow
p_\parallel + 2 p_\perp\leq \varepsilon.
\end{equation}

Let's further suppose a barotropic equation of state:
\begin{equation}\label{I.6}
p_\parallel=k_\parallel \varepsilon;\quad p_\perp=k_\perp \varepsilon,
\end{equation}
where coefficients of baratrops $k_\parallel,\; k_\perp$ by reason
of (\ref{I.5}) are follow the inequality:
\begin{equation}\label{I.7}
k_\parallel+2 k_\perp\leq 1.
\end{equation}

Then in a presence of PGW the exact solution of RMHD equations is
\cite{B5}:
\begin{eqnarray} \label{I.8}
v_2 = 0;\quad v_u=\frac{1}{2v_v};\\
\label{I.9} \displaystyle v_v = \frac{1}{\sqrt{2}} \left[\Delta
L^{k_\parallel
+ k_\perp} e^{\beta(k_\parallel - k_\perp)} \right]^{g_\perp};\\
\label{I.10} \displaystyle \varepsilon =
\stackrel{0}{\varepsilon}\left[\Delta^{1 + k_\perp} L^{ 2 (1 +
k_\parallel)}
 e^{2\beta(k_\parallel - k_\perp)}\right]^{- g_\perp};
\\
\label{I.11} \displaystyle H = H_0 \left[ \Delta L^{ (1 +
k_\parallel)} e^{ - \beta(1 - k_\parallel)} \right]^{-
g_\perp};\\
\label{I.11a} n=\frac{1}{\sqrt{2}}\frac{\stackrel{0}{n}}{v_vL^2},
\end{eqnarray}
where
\begin{equation}  \label{I.12}
g_\perp = \frac{1}{1 - k_\perp} \in [1, 2],
\end{equation}
$\Delta(u)$ is the governing function of GMSW.
\begin{equation}  \label{I.13}
\Delta(u) =  \Bigl[ 1 - \alpha^2 (e^{2\beta} - 1)\Bigr],
\end{equation}
$n$ - local charged particle density, $\alpha^2$ - dimensionless
parameter:
\begin{equation}  \label{I.14}
\alpha^2 = \frac{H_0^2}{4\pi (\stackrel{0}{\varepsilon} +
\stackrel{0}{p}_\perp)}.
\end{equation}

Variables from above marked with zero are given in the
absence of PGW.\\
The RDMD equations solution consists of the physical singularity on
the hypersurface $\Sigma_*: u= u_*$:
\begin{equation}  \label{I.15}
\Delta(u_*) =  \Bigl[ 1 - \alpha^2 (e^{2\beta(u_*)} - 1)\Bigr],
\end{equation}
on which the densities of the plasma energy and of the magnetic
field tend to infinity, the dynamic velocity of the plasma as a
whole tends to the velocity of light in the PGW propagation
direction. In this case the ratio of the magnetic field energy
density to the plasma energy tends to infinity. The above
singularity is the GMSW spreading in the PGW propagation direction
at a subluminal velocity. According to Eq.(\ref{I.15}) the
conditions of the singularity arising are
\begin{equation}  \label{I.16}
\beta(u) > 0;
\end{equation}
\begin{equation}  \label{I.17}
\alpha^2 > 1.
\end{equation}

The extremely important fact is that, the singular condition is even
possible in a weak PGW ($|\beta|\ll 1$) on the condition of a highly
magnetized plasma ($\alpha^2 \gg 1$). In this case the singular
condition, according to (\ref{I.15}), arises on the hypersurfaces $u
= u_*$:
\begin{equation}  \label{I.18}
\beta(u_*) = \frac{1}{2\alpha^2}.
\end{equation}

It follows from (\ref{I.8}) - (\ref{I.11}) that, by $\beta > 0$
the plasma moves in the GW  propagation direction $(v^1 =
\frac{1}{\sqrt{2}}(v_u - v_v) > 0$), by $\beta < 0$ - in the
opposite direction.

The singularity was removed by taking into account back influence
of the magnetoactive plasma on a PGW. It leads to effective
absorption of a PGW energy by the plasma and to PGW amplitude
restriction. The simple model of energy balance which describes
that process was constructed in \cite{B3}. Gravitational wave with
metrics (\ref{I.1}) in WKB-as\-sump\-tion\footnote{In this case it
corresponds to the condition of $\rho\gg\lambda=c/\omega$ where
$\rho$ is a space-time curvature radius.} corresponds to the EMT
with one nonzero component:
\begin{equation} \label{I.19}
\stackrel{gw}{T}_{uu} = \frac{1}{4\pi} (\beta')^2 .
\end{equation}

Let $\beta_*(u)$ be a PGW vacuum amplitude and $\beta(u)$ be a PGW
amplitude in consideration of absorption in plasma. In this case the
energobalance equation in the short-wave approximation becomes:

\begin{equation}  \label{I.20}
(\beta'_*)^2 = (\beta')^2 + 4\pi (T_{uu} -
\stackrel{0}{T}_{uu})\,,
\end{equation}
where $T_{ik}$ is total plasma EMT. Under condition of $\alpha^2
\gg 1$ Eq.(\ref{I.8}) may be written in the form:
\begin{equation}  \label{I.21}
{\dot q}_*^2 = {\dot q}^2 + \xi^2 V(q),
\end{equation}
where $q = \beta/\beta_0$ and the dot signifies a derivative in
the dimensionless time variable $s$:
\begin{equation} \label{s}
s = \sqrt{2} \omega u,
\end{equation}
($\omega$ - the PGW frequency), $V(q)$ - potential function which in
a weak PGW becomes:
\begin{equation}  \label{I.22}
V(q) = \Delta^{- 4g_\perp}(q) - 1,
\end{equation}
where $\xi^2$ is so-called {\it the first parameter of GMSW}
\cite{B2}:
\begin{equation}  \label{I.23}
\xi^2 = \frac{H^2_0}{4 \beta^2_0 \omega^2}.
\end{equation}

Eq.(\ref{I.21}) may be treated as an equation with respect to the
variable $q$. On the other hand, (\ref{I.21}) completely coincides
in its form with the energy conservation law of a 1-dimensional
mechanical system described by the canonical variables $\{q(s),
\dot{q}(s)\}$ \cite{B9}, where $V(q)$ is the potential,
$\dot{q}^2$ is its kinetic energy and $\dot{q^2}_* = E_0$ is its
total energy.\\
Let us introduce the new dimensionless parameter:
\begin{equation}  \label{I.24}
\Upsilon = 2\alpha^2\beta_0
\end{equation}
- ({\it the second GMSW parameter}) and rewrite (\ref{I.15}) in a
weak PGW as:
\begin{equation}  \label{I.25}
\Delta(q(s)) =  1 - 2 \alpha^2 \beta_0 q (s)= 1 - \Upsilon q(s).
\end{equation}

It leads from (\ref{I.25}):
\begin{equation}  \label{I.26}
{\dot q} = - \frac{\dot{\Delta}(q)}{\Upsilon}.
\end{equation}

To analyze the system behavior, let us suppose that the moment $s =
0$ corresponds to the front edge of a GW, while\footnote{This
provides zero PGW metrics derivatives at the moment $s=0$, i.e.
$C^1$  class of metric functions.}:
\begin{equation}  \label{I.27}
\beta_*\approx \beta_0(1-\cos(s))\Rightarrow q_*\approx 1-\cos(s).
\end{equation}

According to (\ref{I.25})-(\ref{I.27}) the system starts with
negative value of the governing function derivative and with
function value equal to 1:
\begin{equation}\label{D_0}
\begin{array}{l}
\dot{\Delta}(s)\approx -\Upsilon\sin s \approx -\Upsilon s ;\\
\\
\Delta(s)\approx 1-\Upsilon(1-\cos s)\approx 1-\Upsilon \frac{s^2}{2};\\
\end{array}
 \quad (s\to +0).
\end{equation}

The energobalance equation (\ref{I.21}) according to (\ref{I.22}),
(\ref{I.26}), (\ref{I.27}) becomes:
\begin{equation}  \label{I.28a}
\dot{\Delta}^2 + \xi^2 \Upsilon^2 \Bigl[\Delta^{- 4g_\perp} - 1
\Bigr] = \Upsilon^2 \sin^2(s).
\end{equation}

Solving the Eq.(\ref{I.28a}) with respect to $\dot{\Delta}$ we
obtain:
\begin{equation}  \label{I.28}
\dot{\Delta}=\mp \Upsilon\sqrt{ \sin^2(s)- \xi^2  \Bigl[\Delta^{-
4g_\perp}(s) - 1 \Bigr] }.
\end{equation}

Integrating according to (\ref{D_0}) first of all it's necessary
to take negative branch of the Eq.(\ref{I.28}) but when we reach
the minimum value of the governing function we should change it by
the positive one. From (\ref{I.28}) we obtain the minimum value of
the governing function which is reached by $s=\pi/2$:
\begin{equation}  \label{I.29}
\Delta_{min} = \left( \frac{1}{\xi^2} + 1\right)^{- \gamma_\perp},
\end{equation}
where:
\begin{equation} \label{I.30}
\gamma_\perp = \frac{1}{4 g_\perp} = \frac{1 -
k_\perp}{4}\Rightarrow \frac{1}{8}\leq \gamma_\perp \leq
\frac{1}{4}.
\end{equation}

The maximum accessible density of a magnetic energy is
\begin{equation}  \label{I.31}
\left(\frac{H^2}{8\pi}\right)_{max} = \frac{H_0^2}{8\pi} \sqrt{1 +
\frac{1}{\xi^2}}
\end{equation}
and it does not depend on a plasma equation of state (\ref{I.6}).
Also plasma velocity in the GMSW does not depend on equation of
state. And the maximum plasma energy density without magnetic
field depends on the exponent of plasma anisotropy:
\begin{equation}  \label{I.32}
\varepsilon_{max} = \stackrel{0}{\varepsilon} \left(1 +
\frac{1}{\xi^2}\right)^{\frac{1}{4} (1 + k_\perp)}.
\end{equation}

It is maximum for the ultrarelativistic plasma with zero valuation
of the parallel pressure.\\
Thus, the maximum value of the local response amplitude of a highly
magnetized plasma ($\alpha^2\gg1$) with linear state equations does
not depend on the exponent of plasma anisotropy and its equation of state.\\

\section{Numerical analysis of the energobalance equation in Mathematica}

Eq.(\ref{I.28}) is essentially nonlinear and difficult for
analyzing in spite of its apparent simplicity. Since there is no
possibility to find the exact solution of the energobalance
equation which has important astrophysical applications there is a
need for its numerical analysis. First attempts of numerical
calculation has met significant troubles. Therefore, for the
numeric integration control the analytic researches were made in
\cite{B6}. They revealed that the solution has a plateau form with
minimum value at the point of $\pi/2$ and after this point the
solution becomes instable. Also some numerical solutions of the
energobalance equation in TurboPascal were obtained in the paper.

Comprehensive analysis of the homogeneous magneto\-active plasma
GMSW response on a gravi\-tational wave within a wide range of
plasma and GW parameters was not done in that time due to the
software abilities. Nowadays the abilities of nonlinear
differential equations numeric solution in computer algebra system
(CAS) Mathematica allow to do such researches. However, the direct
use of build-in numeric methods towards the energobalance equation
(\ref{I.28a}) is still impossible in case that the governing
function derivative changes the sign at the point $s=\pi/2$. So
one can not change the step size according to the equation
parameters.

Empirically was established that the second-order implicit Adams
method solves the equation much faster and much correct in
comparison with other explicit and implicit methods. In this
research the procedure for numerical solving of the differential
equation in CAS Mathematica was developed. It adapts the
integration step according to the para\-meters $\xi^2$,
$\Upsilon$. For all that the differential equation is being
solving with negative value of derivative up to the point $\pi/2$
using the second-order implicit Adams method. The value of
function is being taking as the initial value for the positive
equation branch (\ref{I.28}) after derivative changes the sign.
The integration step is being changed and integration method is
changing to Euler method which works better in the instability
region.

The procedure allows to make analysis of (\ref{I.28}) numerical
solutions depending on the first and the second GMSW parameters.
It also allows to construct the model of magneto\-active plasma
res\-ponse on a GW and to calculate plasma's physical
characte\-ristics. Numerical researches with our proce\-dure
completely approve the analytic predictions for the governing
function form. At first, the solution decreases rapidly then comes
to plateau and slowly approaches the point of minimum $\pi/2$ with
the function value close to the (\ref{I.29}). After the point of
minimum an instability evolves rapidly. For all that the governing
function is smooth in the whole interval. In Fig.\ref{ris1} the
results of numerical solution for the energo\-balance equation in
the case of tiny parameter $\xi^2$ and huge parameter $\Upsilon$
are presented. In this case a process of numeric solution is the
most difficult and on the other hand the predicted features of the
solution are visible. The seeming fractures of function by small
$s$ and by $s=\pi/2$ are fake. In fact they disappear by scaling
up.

Numeric analysis of the energobalance equation allows to determine
the fact that the governing function is sufficiently close to the
function $\Delta_0$ in the plateau area (i.e. the small value of
derivative). $\Delta_0$ nullifies a radical value in right hand
side of (\ref{I.28}):
\begin{equation} \label{I.33}
\Delta_0(s) = \left( 1+\frac{\sin(s)^2}{\xi^2}
\right)^{-\gamma_\perp}.
\end{equation}
\vskip 8pt\noindent \refstepcounter{figure}
\epsfig{file=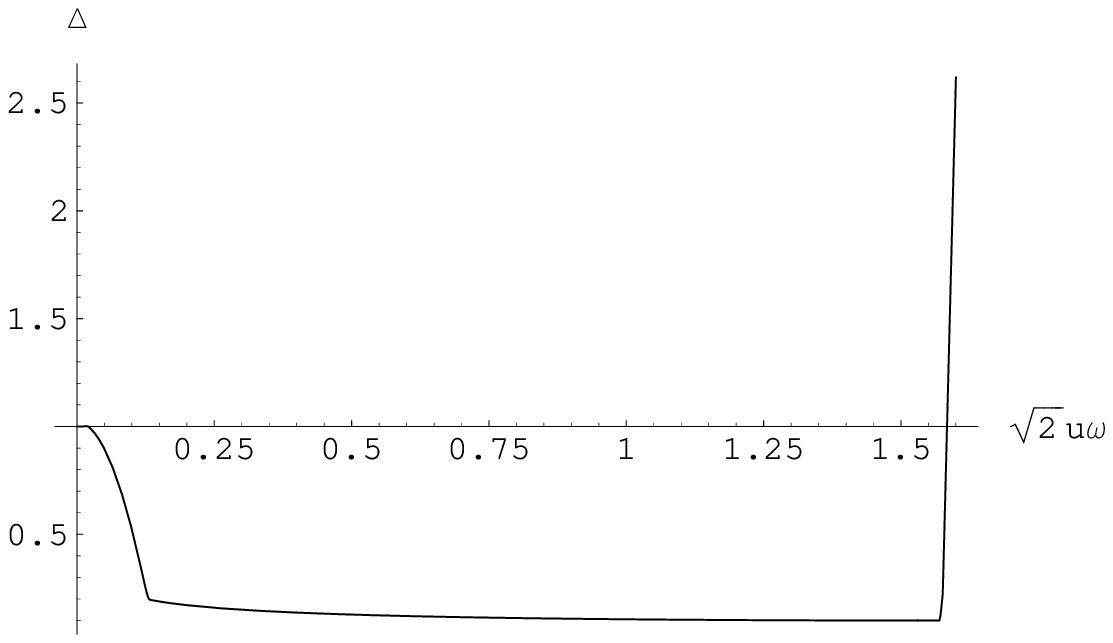,width=8cm}\label{ris1} \vskip 10pt \noindent
{Figure \bf \thefigure.}\hskip 12pt{\sl The governing function
$\Delta(s)$ by $\xi^2=10^{-6}$, $\Upsilon=100$, $\gamma_\perp=1/6$
\hfill}
\vskip 2pt\noindent%

At the point of minimum $s=\pi/2$ this value coincides with the
governing function minimum value (\ref{I.29}). By increasing the
parameter $\Upsilon$ coincidence of the governing function
$\Delta(s)$ and the function $\Delta_0(s)$ becomes by the smaller
values of time $s$. By small values of the time variable $s$ the
governing function is well approximated by parabolic law
(\ref{D_0}). In Fig.\ref{ris2} the plots of the $\Delta(s)$ and
the $\Delta_0(s)$ functions are shown:
\vskip 24pt\noindent \refstepcounter{figure}
\epsfig{file=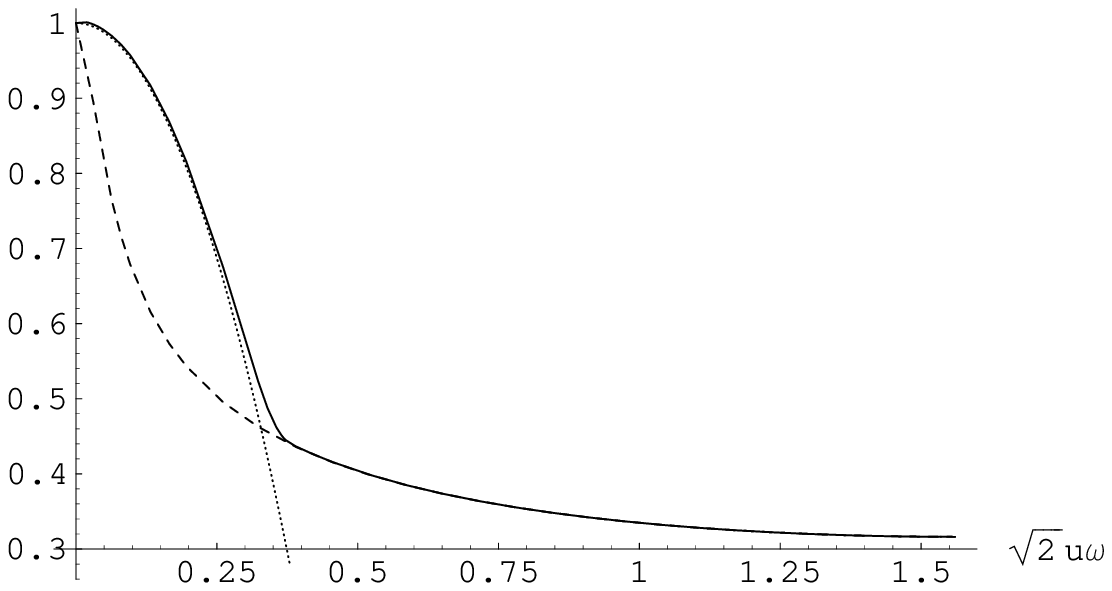,width=8cm}\label{ris2} \vskip 12pt \noindent
{Figure \bf \thefigure.}\hskip 12pt{\sl The $\Delta(s)$ function
(solid line), the $\Delta_0(s)$ function (dashed line), the
asymptotic (\ref{D_0}) by small values of $s$: $1-\Upsilon s^2/2$
(dotted line). Everywhere $\xi^2=0.001$, $\Upsilon=10$,
$\gamma_\perp=1/6$. \hfill}
\vskip 12pt\noindent%

This result allows to approximate the plasma response $H^2/H_0^2$
in the plateau area of the governing function, i.e. in the maximum
response area by the expression:
\begin{equation}\label{Appr_H}
\frac{H^2}{H_0^2}\approx \sqrt{1+\frac{\sin^2s}{\xi^2}}.
\end{equation}
\section{Bremsstrahlung response of homogeneous magnetoactive plasma
on a gravitational wave}

In a weak GW:
\begin{equation}\label{weekgw}
|\beta(s)|\ll 1;\quad L(s)\approx 1
\end{equation}
the exact solution of RMHD equations (\ref{I.8})-(\ref{I.11}) is
simplified and we obtain the following expressions for physical
observed values (see also \cite{B6}).
\vskip 4pt\noindent{\sl 1.}~ Magnetic field energy density in the
comoving frame of reference:
\begin{equation}  \label{I.35}
\frac{H^2}{8\pi} = \frac{H_0^2}{8\pi} \Delta^{- 1/2\gamma_\perp}.
\end{equation}
\noindent{\sl 2.}~ Plasma energy density in the comoving frame of
reference:
\begin{equation}  \label{I.36}
\varepsilon = \varepsilon_0 \Delta^{- \frac{1 + k_{\perp}}{1 -
k_{\perp}}}.
\end{equation}
\noindent{\sl 3.}~ Physical velocity of plasma:
\begin{equation}  \label{I.37}
v^1 = \frac{1 - 2v^2_v}{1 + 2v^2_v} = \frac{1 -
\Delta^{1/2\gamma_\perp}} {1 + \Delta^{1/2\gamma_\perp}}.
\end{equation}
\noindent{\sl 4.}~ Charged particle density:
\begin{equation}\label{I.37a}
n=n_0\Delta^{-1/4\gamma_\perp}.
\end{equation}
\noindent{\sl 5.}~ Total observed bremsstrahlung intensity
detected by resting observer:
\begin{equation} \label{I.38}
W = W_0 \Delta^{-\frac{3+2k_\perp}{1-k\perp}} \frac{1}{2} \left(
\Delta^{1/4\gamma_\perp} + \Delta^{-1/4\gamma_\perp} \right),
\end{equation}
where $W_0$ - total bremsstrahlung intensity in the absence of a PGW
\cite{B8}:
\begin{equation}\label{W_0}
W_0=\frac{2e^4H_0 ^2}{3m^2c^3}n_0\left(\frac{{\mathcal
E}}{mc^2}\right)^2,
\end{equation}
where ${\mathcal E}$ - the kinetic energy of a charged particle.
\noindent{\sl 6.}~  Radiation spectral intensity in a high
frequencies range where frequencies are comparable with the
unperturbed cyclotron frequency $\omega_c^0$:
\begin{equation}\label{omega_c}
\omega_{c}^0 = \frac{3 e H_0}{2 m c}\left( \frac{{\mathcal
E}_0}{mc^2} \right)^2,
\end{equation}
and higher. One may calculate the intensity using standard
electrodynamical formulas \cite{B8} and find:
\begin{equation}\label{J}
J = J_0 \Delta^{-3}
F\left(\frac{\omega}{\omega_{c}^{0}}\Delta^{-5/2}\right),
\end{equation}
where
\begin{equation}
J_0 = \frac{\sqrt{3}}{2\pi}\frac{e^3H_0n_0}{mc^2},
\end{equation}
\begin{equation}\label{F}
F(x) = x \int\limits_{x}^\infty K_{5/3}(y)dy,
\end{equation}
$K_\nu(z)$ -  modified Bessel functions of the third kind or
Macdonald functions (see Ref.\cite{Lebed}):
$$K_\nu(z)=
\frac{\sqrt{\pi}z^\nu}{2^\nu \Gamma(\nu+1/2)}\int\limits_0^\infty e^{-z\cosh t}\sinh^{2\nu}t\  dt. $$

In Fig. \ref{ris4}-\ref{ris3} the results of a numeric solution
for response of magnetoactive plasma on a gravitational wave
depending on GMSW parameters are presented.
\vskip 8pt\noindent \refstepcounter{figure}
\epsfig{file=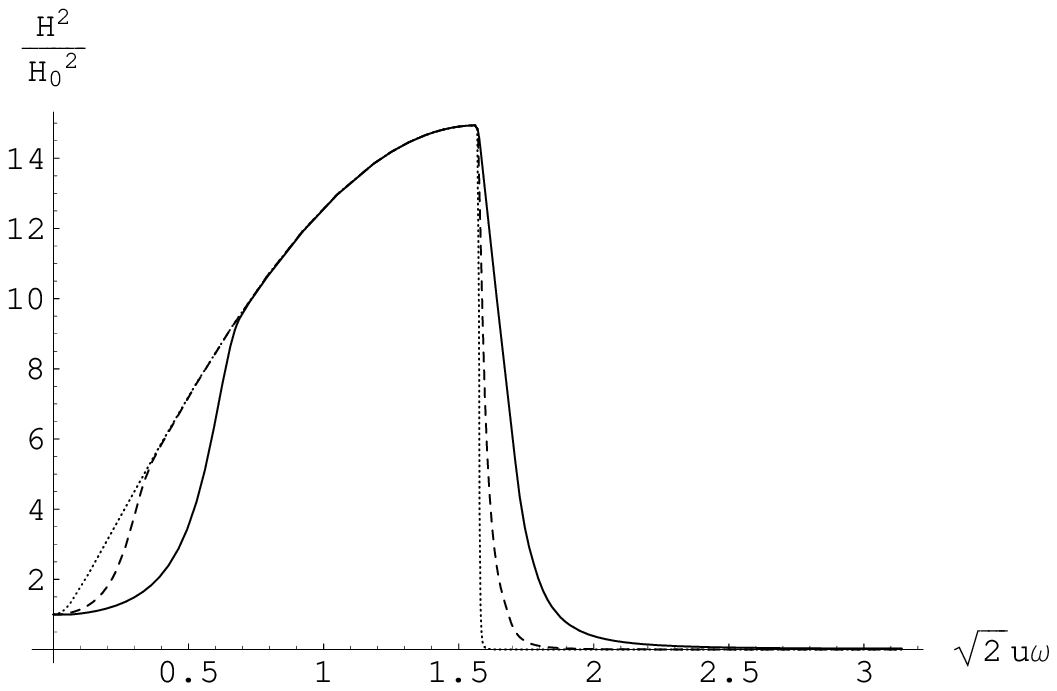,width=8cm}\label{ris4} \vskip 8pt \noindent
{Figure \bf \thefigure.}\hskip 12pt{\sl Influence of the second
GMSW parameter $\Upsilon$ on the relative magnetic field energy
density evolution $H^2/H^2_0$ by $\xi^2=0.0045$,
$\gamma_\perp=1/6$: $\Upsilon=3$ (solid line), $\Upsilon=10$
(dashed line), $\Upsilon=100$ (dotted line). \hfill}
\noindent%
\vskip 8pt\noindent \refstepcounter{figure}
\epsfig{file=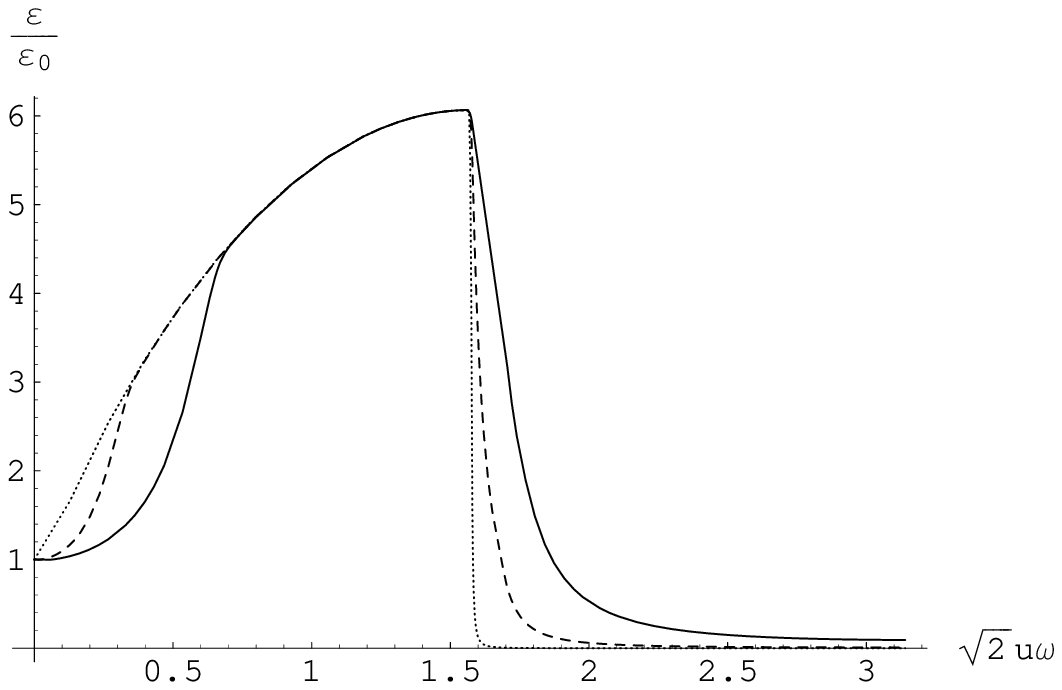,width=8cm}\label{ris5} \vskip 8pt \noindent
{Figure \bf \thefigure.}\hskip 12pt{\sl Influence of the second
GMSW parameter $\Upsilon$ on the plasma energy density evolution
$\varepsilon/\varepsilon_0$ by $\xi^2=0.0045$, $\gamma_\perp=1/6$:
$\Upsilon=3$ (solid line), $\Upsilon=10$ (dashed line),
$\Upsilon=100$ (dotted line). \hfill}
\noindent%
\vskip 8pt\noindent \refstepcounter{figure}
\epsfig{file=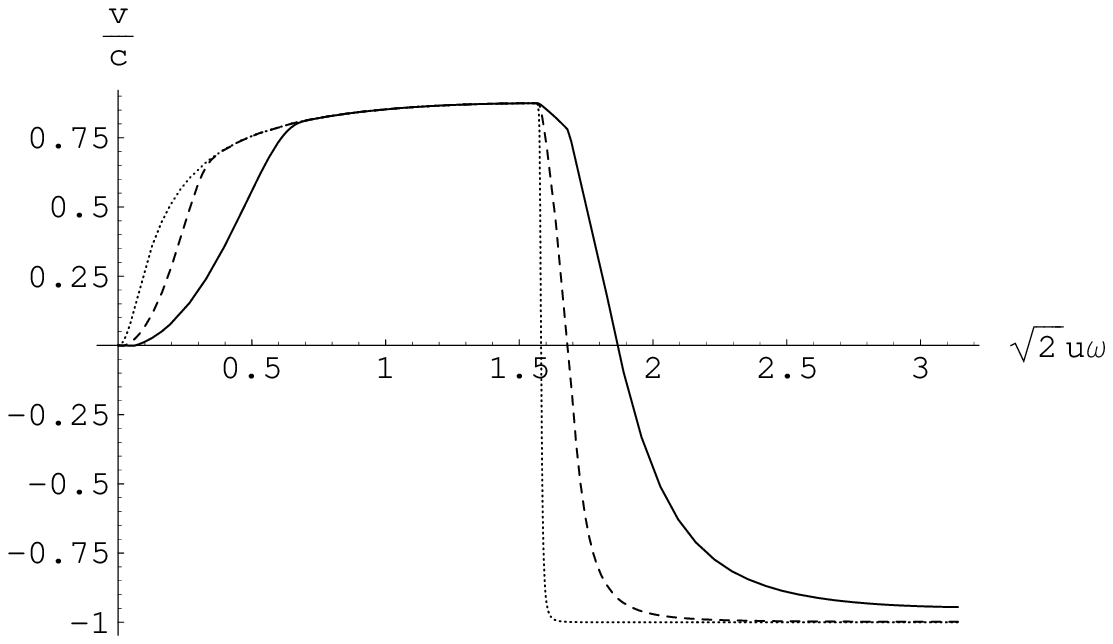,width=8cm}\label{ris6} \vskip 8pt \noindent
{Figure \bf \thefigure.}\hskip 12pt{\sl Influence of the second
GMSW parameter $\Upsilon$ on the plasma drift velocity evolution
$v^1/c$ by $\xi^2=0.0045$, $\gamma_\perp=1/6$: $\Upsilon=3$ (solid
line), $\Upsilon=10$ (dashed line), $\Upsilon=100$ (dotted line).
\hfill}
\vskip 8pt\noindent \refstepcounter{figure}
\epsfig{file=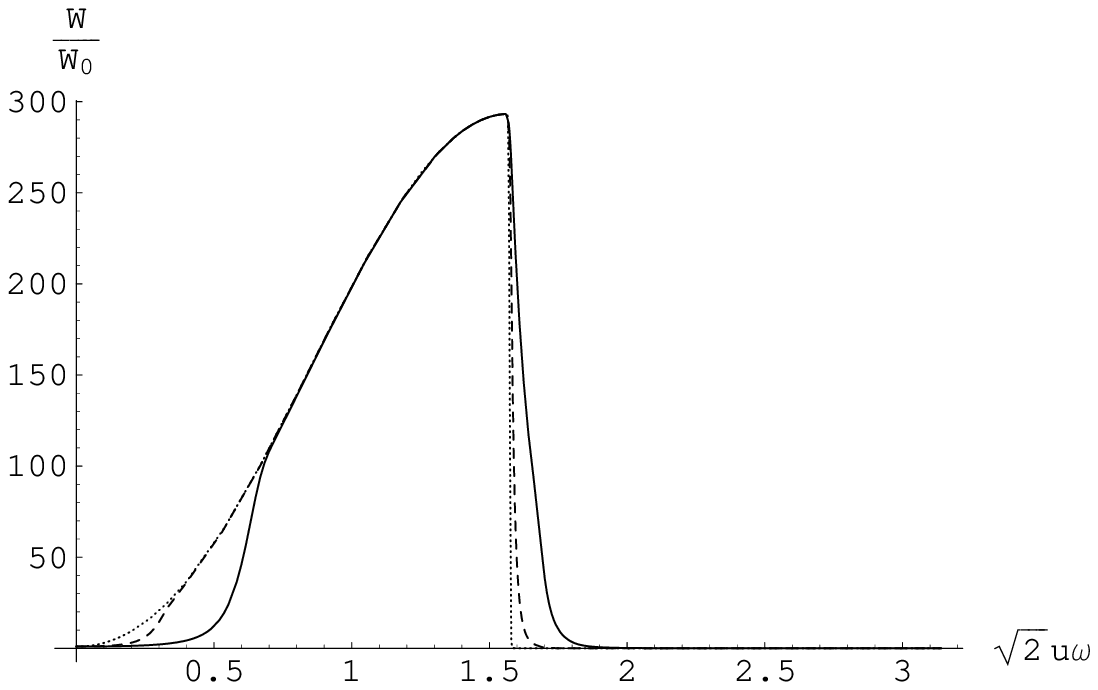,width=8cm}\label{ris7} \vskip 8pt \noindent
{Figure \bf \thefigure.}\hskip 12pt{\sl Influence of the second
GMSW parameter $\Upsilon$ on the total observed bremsstrahlung
intensity evolution $W/W_0$ by $\xi^2=0.0045$, $\gamma_\perp=1/6$:
$\Upsilon=3$ (solid line), $\Upsilon=10$ (dashed line),
$\Upsilon=100$ (dotted line). \hfill}
\vskip 8pt\noindent%

Numeric results allow to determine the region of the parameters
$\xi^2$ and $\Upsilon$ where GMSW mechanism becomes sufficiently
effective. At that the effect was treated as essential if the
total observed brems\-strah\-lung intensity exceeds in maximum its
initial value in about 2 times. As a result of essential
dependence of the total observed brems\-strah\-lung intensity on
the parameter $\xi^2$ this region is close to ellipse
(Fig.\ref{ris3}).
\vskip 24pt\noindent
\refstepcounter{figure}\epsfig{file=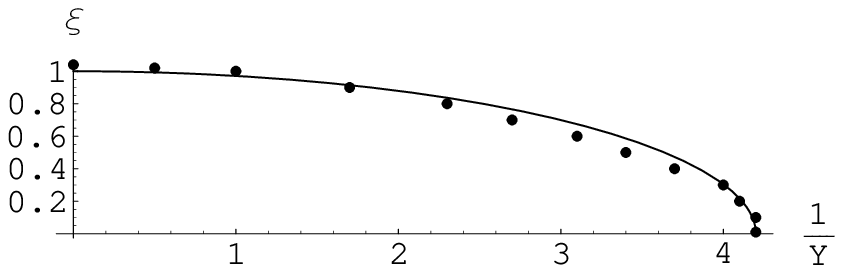,width=8cm} \label{ris3}
\vskip 12pt \noindent {Figure \bf \thefigure.}\hskip 12pt{\sl
Region of the GMSW existence against a quarter ellipse with
semiaxis $4.2$ and $1$ background. \hfill}
\vskip 12pt\noindent%
\vskip 24pt\noindent \refstepcounter{figure}
\epsfig{file=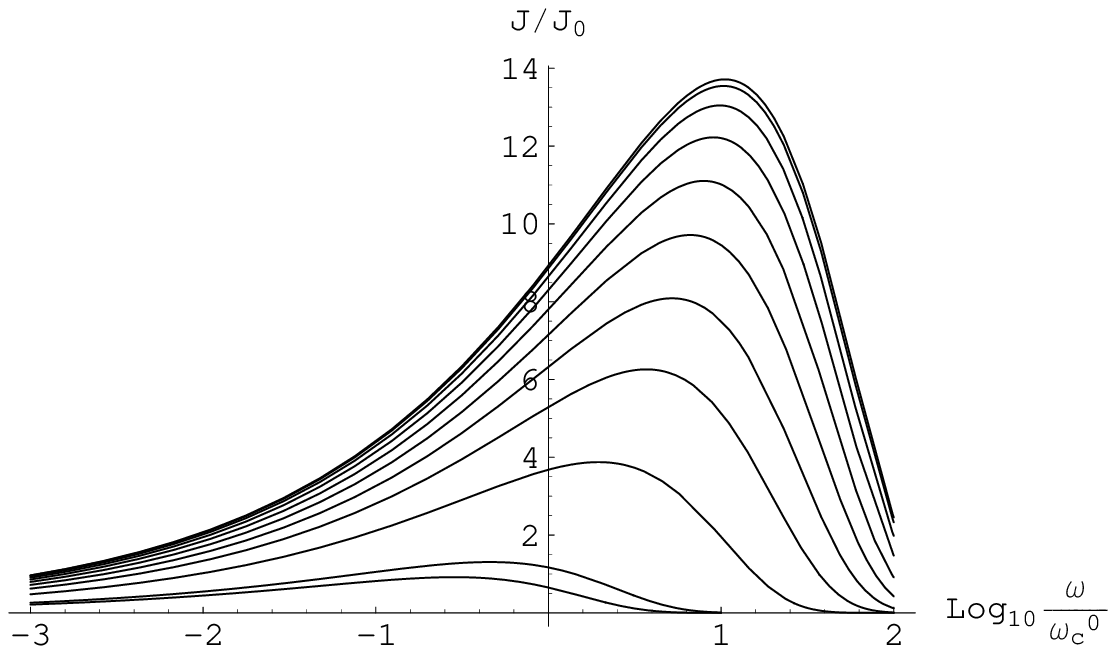,width=8cm}\label{ris8} \vskip 12pt \noindent
{Figure \bf \thefigure.}\hskip 12pt{\sl Bremsstrahlung spectral
intensity time evolution in relative units by $\xi^2=0.0045$,
$\Upsilon=10$ and relative time: $s$=0; 0.15; 0.31; 0.47; 0.63;
0.78; 0.94; 1.10; 1.25; 1.41; 1.56$\approx\pi/2$ (from bottom to
top). \hfill}
\vskip 12pt\noindent%

In Fig.\ref{ris8} along x-coordinate is a common logarithm of
bremsstrahlung frequency in units of the un\-per\-turbed cyclotron
frequency $\omega_{c}^{0}$ and along y-coordinate is relative
radiation intensity $J/J_0$. Maxi\-mum of spectral intensity
(\ref{J}) shifts according to the law:
\begin{equation*}
\omega_{max} = 0,29 \omega_{c}^{0} \Delta^{-5/2}(s).
\end{equation*}
As was mentioned before an instability rapidly evolves when the
governing function passes through its minimum which corresponds to
the observed values maximum. Plasma makes irreversible revers in
the direction opposite to the PGW propagation direction. This
situation is clearly illustrated in Fig.\ref{ris6}. Thus, the
magnetoactive plasma reacts to a PGW by a single impulse
Ref.\cite{B2}, \cite{B3}.
\vskip 24pt\noindent \refstepcounter{figure}
\epsfig{file=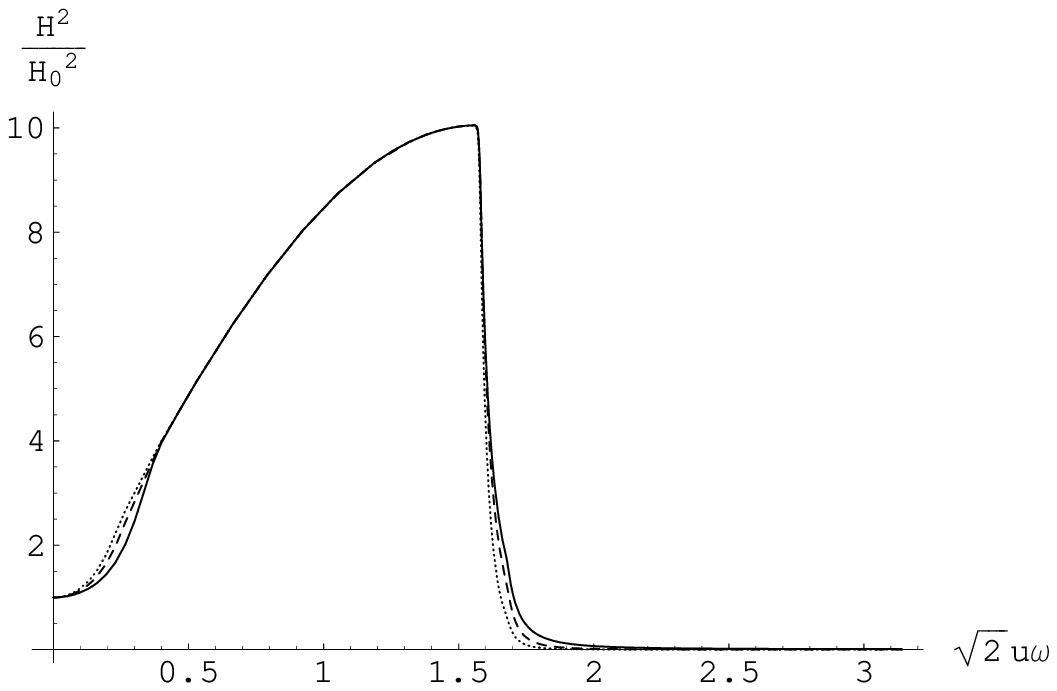,width=8cm}\label{ris9} \vskip 12pt \noindent
{Figure \bf \thefigure.}\hskip 12pt{\sl Influence of the
anisotropy parameter $\gamma_\perp$ on the relative magnetic field
energy density evolution $H^2/H^2_0$ by $\xi^2=0.01$,
$\Upsilon=10$: $\gamma_\perp=1/4$ (solid line), $\gamma_\perp=1/6$
(dashed line), $\gamma_\perp=1/8$ (dotted line). \hfill}
\vskip 12pt\noindent%

The numeric analysis results of the anisotropy parameter
$\gamma_\perp$ influence on the observed magnetic field energy
density are presented in Fig.\ref{ris9}. One can see that
resultant anisotropy factor influence is insignificant in spite of
the essential dependence of the exact solution (\ref{I.8}) -
(\ref{I.12}) on this factor.

The dependence of the total observed brems\-strah\-lung intensity
semiwidth on the GMSW para\-meters was researched.
One can see in Fig.\ref{ris10} that the GMSW impulse duration i.e.
the impulse semi\-width to a high accuracy is equal to
$\pi/4\approx0.79$ or in common units:
\begin{equation}  \label{39}
\delta \tau = \frac{T}{8},
\end{equation}
where $T$ is a PGW period.
\vskip 12pt\noindent \refstepcounter{figure}
\epsfig{file=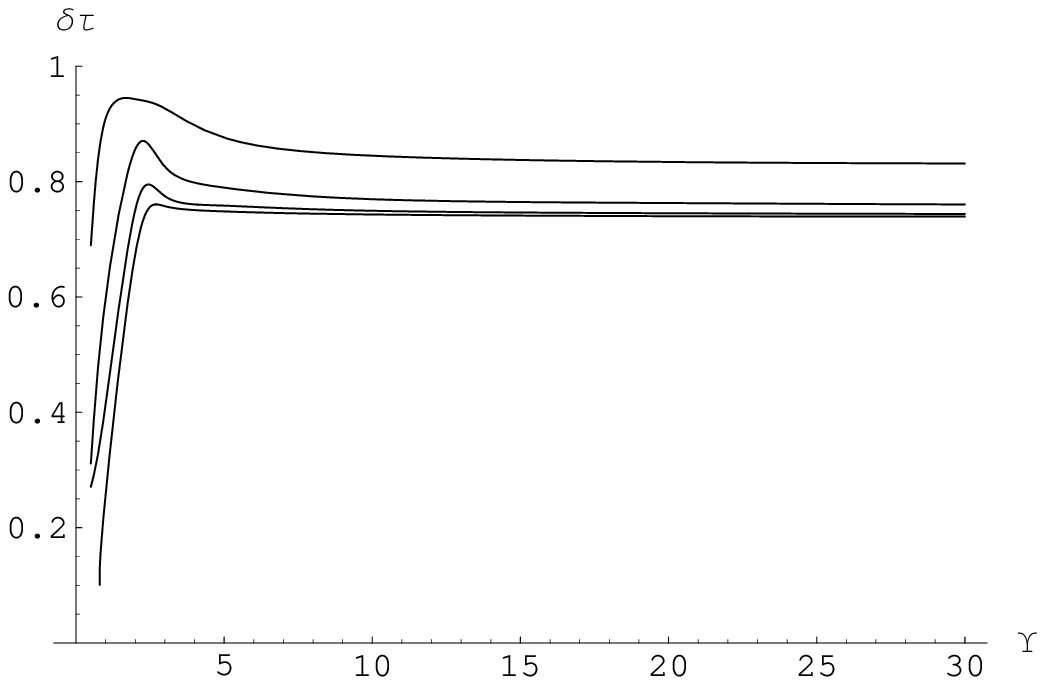,width=8cm}\label{ris10} \vskip 12pt \noindent
{Figure \bf \thefigure.}\hskip 12pt{\sl Dependence of the total
observed brems\-strah\-lung  intensity semi\-width $W/W_0$ on
$\Upsilon$ by $\gamma_\perp=1/6$, $\xi^2=0.1$, $\xi^2=0.01$,
$\xi^2=0.001$, $\xi^2=0.0001$ (from top to bottom). \hfill}
\vskip 12pt\noindent%

\section{Conclusion}

Summarizing the paper results we would like to underline that
gravitational wave weakness is con\-sidered in terms of the
conditions (\ref{weekgw}) realization. In this case the value of
$\alpha^2\beta$ may not be small. Therefore linearity of the
theory by GW smallness in comparison with 1 (linearity of the
Einstein equations left hand side by $\beta$) in general does not
mean linearity of hydrodynamic theory by GW amplitude smallness.
Let us notice that such situation is rather unexpected though it
can be foreseen using a MHD equations exact dimensional analysis.

The research proved preliminary results of the earlier papers and
helped to work out in details the GMSW behavior and to describe
its evolution process in all regions of the parameters.

Numeric simulation of a magnetoactive plasma response on a
gravitational wave allows to find next rules of the gravimagnetic
shock wave excitation process:\\
\noindent{\sl 1.}~ Under realization of the GMSW origin conditions
\begin{equation}\label{Upxi}
\xi<1;\quad \Upsilon>1
\end{equation}
magnetoactive plasma reacts to a PGW by a single impulse where
plasma moves in the gravitational wave propagation direction. The
impulse semiwidth order is $1/8$ of GW period;\\
\noindent{\sl 2.}~ Impulse stops with plasma revers; by this
appears the typical impulse form (see Fig.\ref{ris11}). It weakly
depends on the second GMSW parameter $\Upsilon$ and is determined
by the $\Delta_0(s)$ function.
\vskip 24pt\noindent \refstepcounter{figure}
\epsfig{file=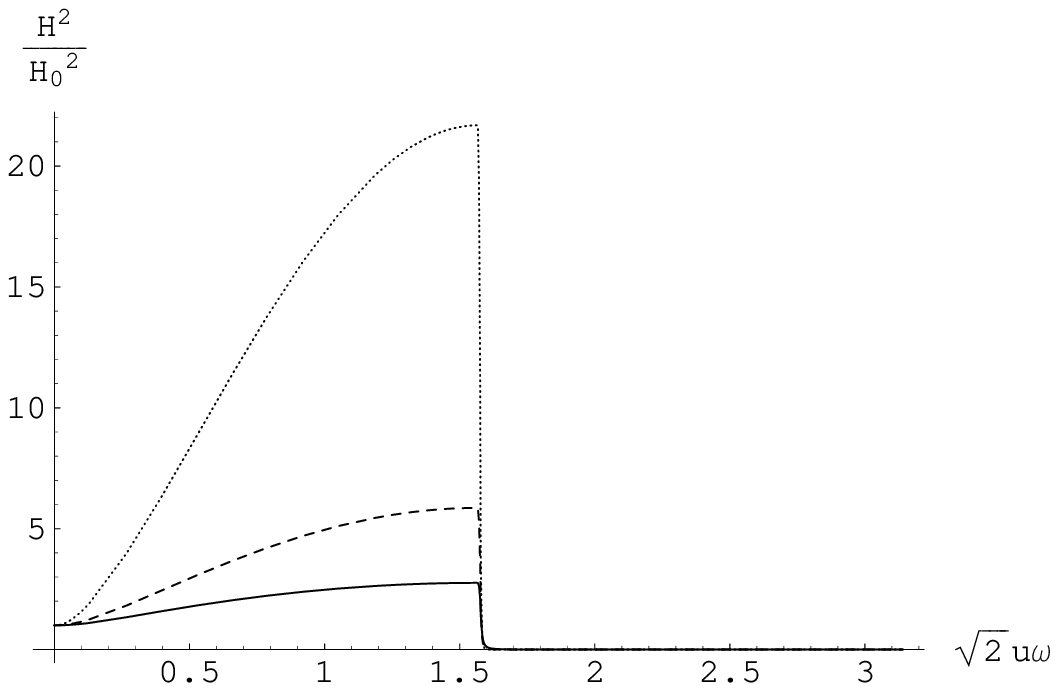,width=8cm}\label{ris11} \vskip 12pt \noindent
{Figure \bf \thefigure.}\hskip 5pt{\sl Influence of the first GMSW
parameter on the magnetic field energy density evolution
$H^2/H^2_0$ by $\Upsilon=100$, $\gamma_\perp=1/6$:~ $\xi^2=0.5$
(solid line), $\xi^2=0.3$ (dashed line), $\xi^2=0.1$ (dotted
line). \hfill}
\vskip 12pt\noindent%
Under conditions (\ref{Upxi}) the second GMSW parameter $\Upsilon$
affects only on front (small values of time $s$) and back (time
values $s$ are close to $\pi/2$) edges of the
impulse.\\
\noindent{\sl 3.}~ Bremsstrahlung spectrum becomes harder during a
shock wave passing.
\\
\noindent{\sl 4.}~ In the maximum of magnetoactive plasma response
almost all of a gravitational wave energy transfers to plasma,
magnetic field and to brems\-strah\-lung (see Fig.\ref{ris12}).

\subsection*{Acknowledgement}
The authors are grateful to Prof. D.V.~Galtsov, B.E.~Meie\-ro\-vich and
N.I.~Kolosnitsin for useful discussion of the results.\\
The authors are grateful to Dr. A.V.~Matrosov, who pointed out the stiff class
of the energobalance equation.

\vskip 24pt\noindent \refstepcounter{figure}
\epsfig{file=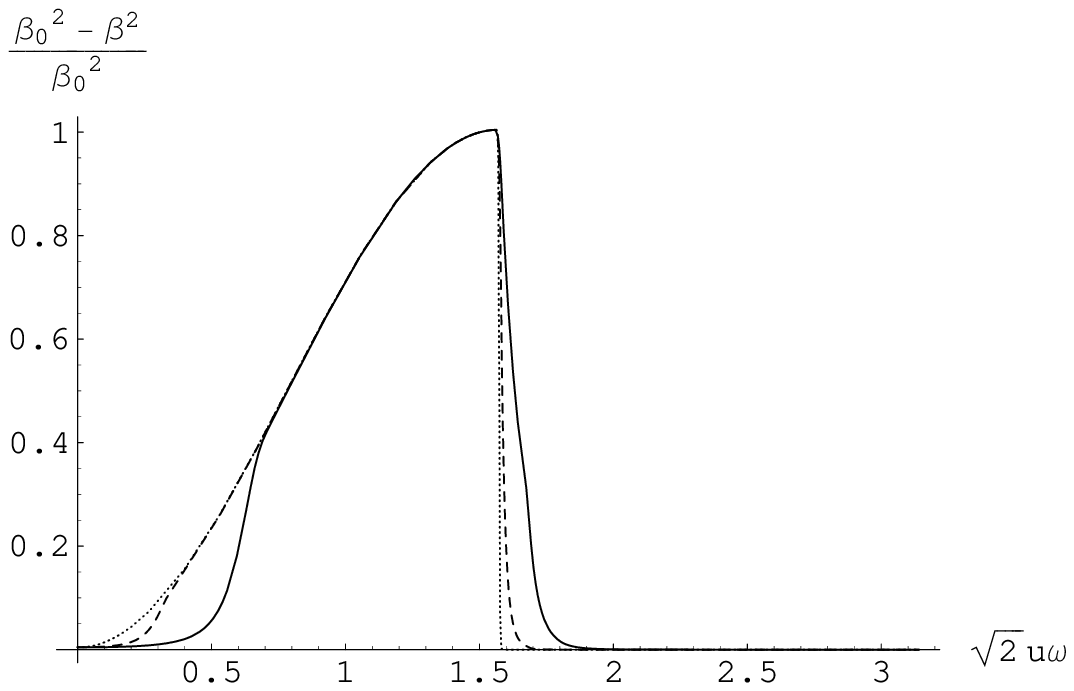,width=8cm}\label{ris12} \vskip 12pt \noindent
{Figure \bf \thefigure.}\hskip 12pt{\sl Influence of the second
GMSW parameter $\Upsilon$ on the GW energy absorbtion
$\Delta\varepsilon_g/\varepsilon_g=(\beta^2_0-\beta^2)/\beta^2_0$
by $\xi^2=0.0045$, $\gamma_\perp=1/6$: $\Upsilon=3$  (solid line),
$\Upsilon=10$ (dashed line), $\Upsilon=100$ (dotted line). \hfill}
\vskip 12pt\noindent%

\end{document}